\documentclass[10pt,twocolumn]{article}
\usepackage{graphicx}
\usepackage{amsmath,amssymb,times,cite}
\usepackage{color,array}
\usepackage{booktabs}
\usepackage{graphicx}
\usepackage{amsmath} 
\usepackage{multirow}
\usepackage{url}
\usepackage{footnotebackref}

\usepackage{authblk}

\usepackage[square,numbers,sort&compress]{natbib}

\usepackage{algorithm,algorithmic}
\makeatletter
\newcommand{\ALOOP}[1]{\ALC@it\algorithmicloop\ #1%
  \begin{ALC@loop}}
\newcommand{\ENDALOOP}{\end{ALC@loop}\ALC@it\algorithmicendloop}

\makeatother
\usepackage{multirow}
\usepackage{booktabs}

\usepackage[table]{xcolor}

\definecolor{lightgray}{gray}{0.9}
\definecolor{lightblue}{rgb}{0.93,0.95,1.0}
\usepackage{authblk}
\usepackage{hyperref}

\title{\Large\textbf{Enhanced LLM-Based Framework for Predicting Null Pointer Dereference in Source Code.}}

\author{
    Md. Fahim Sultan, Tasmin Karim, Md. Shazzad Hossain Shaon, Mohammad Wardat, Mst Shapna Akter\\
    Department of Computer Science and Engineering, Oakland University,\\ Rochester, MI 48309, USA. \\
     mdfahimsultan@oakland.edu, tasminkarim@oakland.edu,\\ shaon@oakland.edu, wardat@oakland.edu, akter@oakland.edu
}

\date{}

\begin{document}
\maketitle

\begin{abstract}
Software security is crucial in any field where breaches can exploit sensitive data, and lead to financial losses. As a result, vulnerability detection becomes an essential part of the software development process. One of the key steps in maintaining software integrity is identifying vulnerabilities in the source code before deployment. A security breach like CWE-476, which stands for NULL pointer dereferences (NPD), is crucial because it can cause software crashes, unpredictable behavior, and security vulnerabilities. In this scientific era, there are several vulnerability checkers, where, previous tools often fall short in analyzing specific feature connections of the source code, which weakens the tools in real-world scenarios. In this study, we propose another novel approach using a fine-tuned Large Language Model (LLM) termed "DeLLNeuN". This model leverages the advantage of various layers to reduce both overfitting and non-linearity, enhancing its performance and reliability. Additionally, this method provides dropout and dimensionality reduction to help streamline the model, making it faster and more efficient. Our model showed 87\% accuracy with 88\% precision using the Draper VDISC dataset. As software becomes more complex and cyber threats continuously evolve, the need for proactive security measures will keep growing. In this particular case, the proposed model looks promising to use as an early vulnerability checker in software development.

\end{abstract}

\bigskip

\noindent\textbf{Keywords}:\, Software Security; Vulnerability Detection; NULL Pointer Dereference; Cybersecurity; CWE-476; LLM.

\vspace{3em}

\section{Introduction}
  In an increasingly interconnected digital world, businesses and customers become more integrated, it is now crucial to ensure software system security before these breaches lead to significant financial losses, damage to reputation, and legal issues \cite{ref1}. Breaches could arise only because of vulnerabilities. Vulnerability is referred to as a flaw or gap in the code that, if not fixed, can be exploited by a threat source \cite{ref2}. These vulnerabilities occur due to various factors, such as poor coding practices, lack of input validation, or improper use of system resources. There are a variety of vulnerabilities in code analysis termed Common Weakness Enumeration (CWE) such as CWE-119 (Buffer errors), CWE-120 (Buffer overflow), CWE-469 (Pointer miscalculation), CWE-476 (NULL pointers), etc. In this study, we focused on efficiently finding the vulnerabilities of CWE-476 from the source code. CWE-476 describes the issue of NPD, which occurs when an application tries to use a pointer that it assumes is valid but is NULL. According to the study, NPD has been listed in the “CWE Top 25 Most Dangerous Software Weaknesses” \cite{ref3} on the CWE website since 2019. Table.~\ref{tab:tab1} displays the CWE Top 25 from the CWE website as of 2023. There could be several causes of NPD.

NULL dereferences can arise in various ways across different CVEs (Common Vulnerabilities and Exposure). For instance, handling an excessive number of packets might lead to a NULL dereference, revealing potential flaws in how packets are managed. Similarly, improper memory initialization can also result in a NULL dereference. Another common issue is unchecked return values, which can lead to NULL dereferences if not properly handled. These are just a few examples of how NULL dereferences can occur. As a result, NULL dereferences often cause crashes or unexpected exits, making the application vulnerable to exploitation and instability. 

In the past, security issues were mostly seen as technical problems, with a focus on finding tech solutions. However, after many studies, it's now clear that information security needs to be addressed from a management perspective as well \cite{ref4}. During development, it's crucial to fix any security flaws in both the source code and technology before deploying the software. However, identifying CWE-476 in source code can be challenging for developers. It requires a thorough analysis of all code paths, especially in large or complex projects. Pointers might be initialized under certain conditions that are hard to trace. Additionally, manual code reviews can also be time-consuming and prone to errors. Thus, a vulnerability checker is a valuable tool for developers, helping to efficiently identify and address security gaps in the code \cite{ref5}.

LLMs have revolutionized the world of artificial intelligence (AI), bringing powerful capabilities to tasks like natural language processing, machine translation, and question-answering \cite{ref6}. With their deep understanding of human language, LLMs are making modern technology more adaptive and effective. They allow recommendation systems to better understand and infer relationships between user features, behavioral patterns, and the connections among entities \cite{ref7}. In this study, we explored a novel method DeLLNeuN (Vulnerability \textbf{De}tection by leveraging \textbf{LL}M and \textbf{Neu}ral \textbf{N}etwork) that combines data-driven techniques with fine-tuned LLMs to predict vulnerabilities effectively. We applied our methodology and conducted a set of experiments using various subsets of the Draper VDISC Dataset \cite{ref8, ref9}. We used the pre-trained LLM model as a feature extractor and built a custom neural network layer that provided reliable performance. This approach has shown promising results in identifying vulnerabilities. 

The main contributions of this article include:
\begin{itemize}
    \item Leverage LLMs to deeply understand and analyze the code by capturing its detailed syntactic and semantic features.
    \item The study developed a robust classifier based on neural network, where, incorporating dropout, a dense layer, and ReLU activation, which are essential for enhance accuracy of vulnerability detection.
    \item Evaluate the model using various evaluation metrics and deep analysis to provide a comprehensive assessment of its effectiveness.
\end{itemize}

To the best of our knowledge, relatively few studies have applied  deep learning techniques to directly learn features from source code in a large natural codebase for detecting vulnerabilities. We believe this approach offers a promising alternative to previous methods, showing great potential for high performance.\vspace{1em}
 
\textbf{Paper organization} The paper is organized in different sections for the future developers and readers such as Section \S 2 reviews related work, while Section \S 3 details the materials and methods used in this study. Section \S 4 presents the experimental results, followed by Section \S 5, which discusses limitations and potential directions for future research. Finally, Section \S 6 provides the conclusion.

\section{Related Works}

  According to the previous outcome, lots of researchers wanted to mitigate the challenges of vulnerabilities such as Lu et al. \cite{ref10} recently introduced a tool named GRACE, which improves LLM by adding graph structure information. This enhancement helps LLMs grasp the interconnected context within source code more effectively. However, GRACE faces several challenges: it becomes computationally intensive when dealing with large codebases due to the increased complexity of managing the graph, and it struggles with the diverse coding styles in C/C++ code, potentially leading to missed vulnerabilities. 

Another study, Wang et al. \cite{ref11} developed DefectHunter model based on combination of convolutional networks and self-attention mechanisms to accurately identify software vulnerabilities. However, the model have some challenges with computation complexity, particularly with large codebases, and struggles with diverse coding styles, which can lead to misidentifying vulnerabilities. Additionally, Akuthota et al. \cite{ref12} introduced a pre-trained GPT-3.5-Turbo model, where its provided 77\% of accuracy, which is an insufficient rate for the future.

Bilgin et al. \cite{ref13} used a public dataset that contains inconsistently labeled data, which is not always accurate. Additionally, the vectorial representations of source code, particularly those based on partial Abstract Syntax Trees (ASTs), can lose important contextual information. Consequently, the machine learning model they developed could not generalize to programming languages. Some studies, like Shin et al. \cite{ref14}, and Alon et al. \cite{ref15} have explored whether software metrics such as code complexity, code changes, and developer activity, which helps to predict the vulnerable code areas. The authors have evaluated with some metrics on open-source projects like Mozilla Firefox and Red Hat Enterprise Linux.
Zhen et al. \cite{ref16} proposed a comprehensive approach for vulnerability detection that integrates Bidirectional Recurrent Neural Networks (BRNN-vdl) with initial analysis using Lower Level Virtual Machine (LLVM) intermediate representations. In VulDeeLocator, BRNN-vdl is employed to refine and narrow down the code. Concurrently, LLVM Intermediate Representation (IR) is utilized to simplify the original source code into a low-level format. However, narrowing down the code can improve detection accuracy, it may still fail to identify the most precise location of the vulnerability. Additionally, relying on LLVM for code analysis is not universally applicable, as it is not suitable for all programming languages. Chernis et al. \cite{ref17} showed that simple metrics, such as character diversity and function length, can also be useful for spotting potential vulnerabilities. However, Walden et al. \cite{ref18} found that these software metrics alone were insufficient for effectively detecting vulnerabilities. In response, the authors introduced an approach that treats source code as regular text and applies natural language processing techniques for code representation and feature extraction. Building on this idea, Russell et al. \cite{ref8} developed an NLP-based sentiment analysis model to classify vulnerabilities within code with employing deep feature representation learning. Nevertheless, the ever-changing structure of code poses challenges for embeddings used in NLP, as the authors struggle to consistently capture code semantic details. 

Alon et al. \cite{ref19} employed a model that represent code snippets as continuous vector which capture the relationship between code snippets and its labels. However, the approach lacks effectiveness in capturing vulnerabilities within dynamic coding environments. Additionally, several studies \cite{ref20, ref21, ref22} have explored alternative methods to improve the efficiency of source code representation in ML-based model analysis, focusing on reducing information loss during the representation learning process by using Abstract Syntax Tree (AST) based encoding. However, ASTs often fall short of capturing the semantic details and broader context needed to identify vulnerabilities, especially in complex and dynamic C/C++ code. This limitation can result in incomplete or inaccurate analysis, making it harder to detect certain vulnerabilities that depend on a deeper understanding of the code's behavior and context.

Therefore, based on these limitations, we have design another approach to detect the CWE-476 detection for future developers with LLM base model. We believe this article will helps and reduce the vulnerabilities with more effectively.

\section{Materials and Methods}

\subsection{Dataset Descriptions}
\begin{figure}
    \centering
    \includegraphics[width=0.99\linewidth]{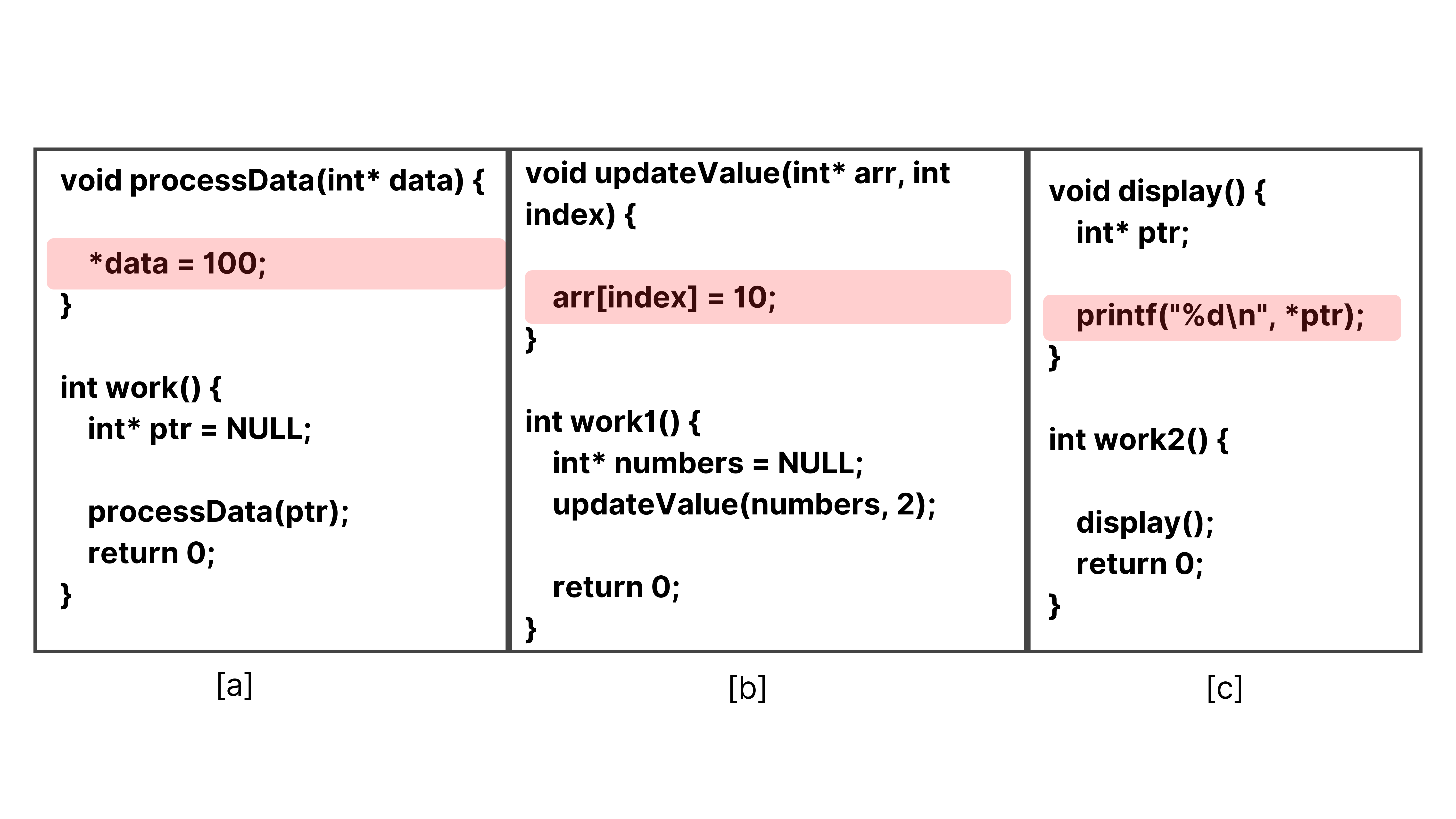}
    \vspace{-2em}
    \caption{An illustration of null pointer dereference (NPD) vulnerabilities in source code. Red markings highlight where NULL pointers are present.}
    \label{fig:f1}
\end{figure}

In our approach to NPD identification, we utilized publicly available datasets that have been extensively used in previous research Table.~\ref{tab:tab2}  shows the overall dataset overview. Initially, we collected the dataset from \url{https://osf.io/d45bw/}, which is publicly accessible. In this study, we collected 24,492 samples, of which 18,880 were set aside for training and 5612 for testing. The following dataset primarily consists of two labels: TRUE, which denotes vulnerability, and FALSE, which denotes non-vulnerability.  
 An illustration of a NPD vulnerability in the source code is displayed in Figure \ref{fig:f1}, with red markings highlighting where NULL pointers are present.

 Figure \ref{fig:f1} is an example of null pointer dereference vulnerability, [a] indicating dereferencing without null check, [b] indicating incorrect pointer arithmetic, [c] indicating uninitialized pointer
In these examples, null pointer dereferences occur because the pointers aren’t properly checked or set up. In Figure \ref{fig:f1}[a], a function tries to use a pointer without making sure it isn’t NULL, which can cause the program to crash if the pointer is `NULL`. In Figure \ref{fig:f1}[b] scenario, an array pointer is used as if it’s valid, but if it’s NULL, trying to access or change the array can lead to unexpected behavior. In Figure \ref{fig:f1}[c], a pointer is used without being set up first, which can result in accessing a bad memory location. These examples show why it’s important to always check and initialize pointers to avoid these issues. Others wise, the system could be vulnerable to exploitation by a threat source.
Our machines are naturally not able to understand categorical or contextual values. We used the pre-trained model CodeBERT as feature embedders to transform these textual values into numerical values and uncover the relevant contextual information.

\begin{table}[htbp]  
\centering
\caption{Distribution of Dataset Labels: TRUE and FALSE.}
\vspace{1em}
\label{tab:dataset_distribution}
\resizebox{0.7\linewidth}{!}{
\begin{tabular}{|l|c|c|c|}
\hline
\textbf{Dataset} & \textbf{TRUE set} & \textbf{FALSE set} & \textbf{Total Set} \\ \hline
Train Set        & 9,440             & 9,440              & 18,880             \\ \hline
Test Set         & 2,810             & 2,802              & 5,612              \\ \hline
\end{tabular}%
}
\label{tab:tab2}
\end{table}

\subsection{ Overview of the Proposed Methodology}

\begin{figure*}
    \centering
    \includegraphics[width=1\linewidth]{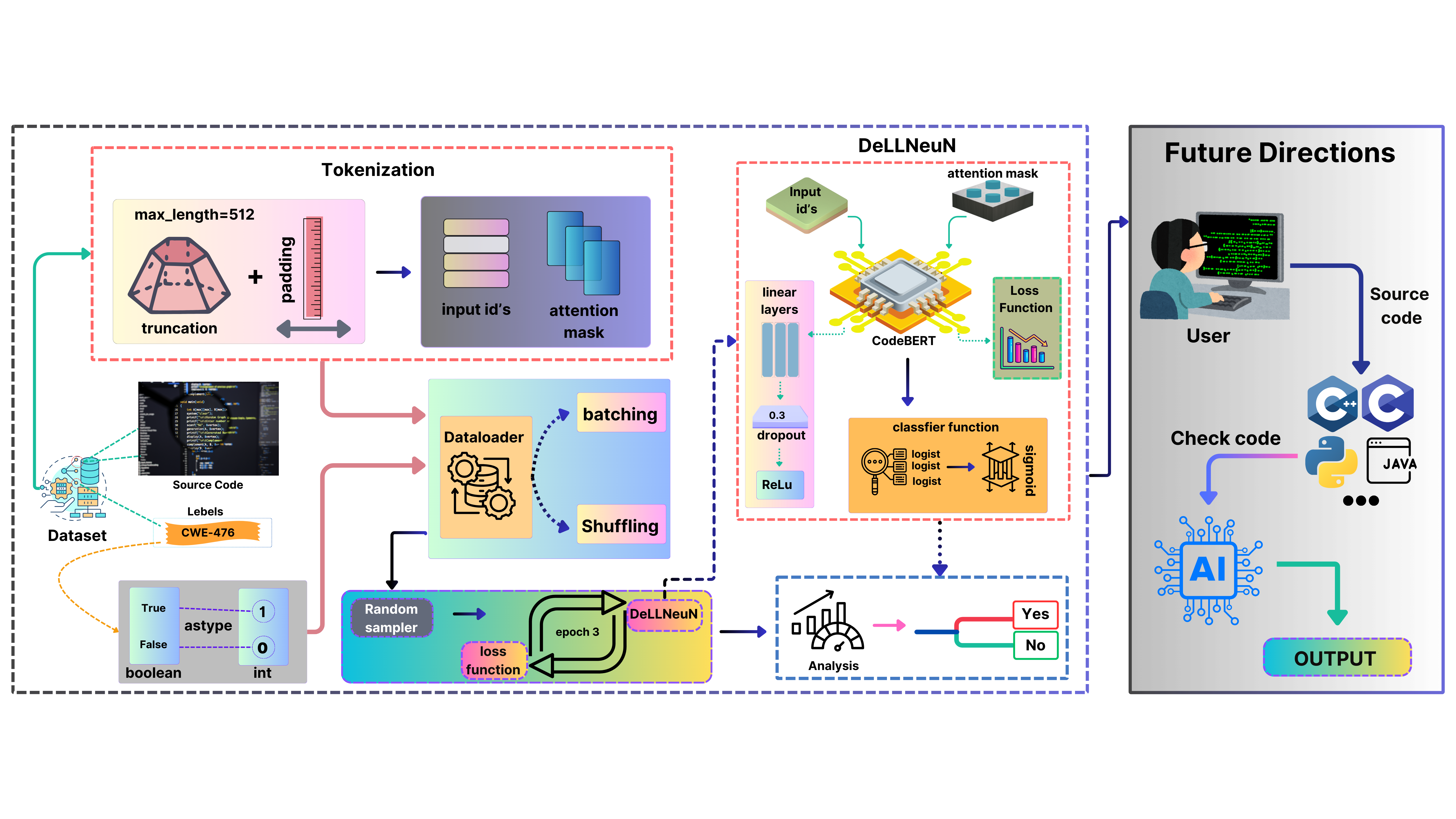}
    \vspace{-4em}
    \caption{Overall methodology of the proposed approach.}
    \label{fig:f22}
\end{figure*}

 Initially, we began by gathering the necessary dataset \cite{ref8} and then performed specific preprocessing steps tailored to our models. We leveraged a pre-trained LLM to uncover the contextual information within our source code segments and transform them into vector representations. Additionally, we performed type casting on the Boolean data in our target class, converting it into numeric values to ensure compatibility with our model, as it is highlighted in Figure \ref{fig:f22}.
 
\definecolor{customblue}{HTML}{78B7D0}

\begin{table}[htbp]
\centering
\caption{CWE Top-25 Most Dangerous Software Weaknesses Statistics in 2023.}
\vspace{1em}
\label{tab:cwe_top_25}
\resizebox{\linewidth}{!}{%
\begin{tabular}{|c|c|p{7cm}|}
\hline
\textbf{Position} & \textbf{Identity} & \textbf{Overview} \\ \hline
1  & CWE-787 & Out-of-bounds Write \\ \hline
2  & CWE-79  & Improper Neutralization of Input During Web Page Generation (Cross-site Scripting) \\ \hline
3  & CWE-89  & Improper Neutralization of Special Elements used in an SQL Command (SQL Injection) \\ \hline
4  & CWE-416 & Use After Free \\ \hline
5  & CWE-78  & Improper Neutralization of Special Elements used in an OS Command (OS Command Injection) \\ \hline
6  & CWE-20  & Improper Input Validation \\ \hline
7  & CWE-22  & Improper Limitation of a Pathname to a Restricted Directory (Path Traversal) \\ \hline
8  & CWE-352 & Cross-Site Request Forgery (CSRF) \\ \hline
9  & CWE-434 & Unrestricted Upload of File with Dangerous Type \\ \hline
10 & CWE-862 & Missing Authorization \\ \hline
\rowcolor{customblue} 
11 & CWE-476 & NULL Pointer Dereference \\ \hline
12 & CWE-287 & Improper Authentication \\ \hline
13 & CWE-19  & Integer Overflow or Wraparound \\ \hline
14 & CWE-502 & Deserialization of Untrusted Data \\ \hline
15 & CWE-77  & Improper Neutralization of Special Elements used in a Command (Command Injection) \\ \hline
16 & CWE-119 & Improper Restriction of Operations within the Bounds of a Memory Buffer \\ \hline
17 & CWE-798 & Use of Hard-coded Credentials \\ \hline
18 & CWE-918 & Server-Side Request Forgery (SSRF) \\ \hline
19 & CWE-306 & Missing Authentication for Critical Function \\ \hline
20 & CWE-362 & Concurrent Execution using Shared Resource with Improper Synchronization (Race Condition) \\ \hline
21 & CWE-269 & Improper Privilege Management \\ \hline
22 & CWE-94  & Improper Control of Generation of Code (Code Injection) \\ \hline
23 & CWE-863 & Incorrect Authorization \\ \hline
24 & CWE-276 & Incorrect Default Permissions \\ \hline
25 & CWE-862 & Missing Authorization \\ \hline
\end{tabular}%
}
\label{tab:tab1}
\end{table}

After that, we fed the preprocessed data into our proposed model, DeLLNeuN, training it over three epochs with a carefully chosen loss function to optimize weight adjustments throughout the process. Our DeLLNeuN model is designed with neural layers that help prevent overfitting, reduce dimensionality, and introduce non-linearity, all of which are crucial for improving the model's performance. 
We evaluated the model's performance based on several key metrics, including accuracy, f1 score, precision, and recall. After conducting thorough analyses and comparisons with other models, we found that DeLLNeuN consistently delivered outstanding performance across all metrics, demonstrating its effectiveness and robustness in vulnerability detection.

\subsection{Dataset Preprocessing}

Initially, we utilize the pre-trained CodeBERT model, which is designed for understanding both code snippets and natural text. CodeBERT is based on the concept of Bidirectional Encoder Representations from Transformers (BERT) \cite{ref21}. In this study, tokenization and encoding are two key steps of the preprocessing pipeline. The `astype` function of pandas is employed for the labels class. The tokenizer generates token IDs and attention masks from text code snippets and ensures that the sequences are standardized by padding and truncating to a maximum length of 512 tokens.

\begin{figure}
    \centering
    \includegraphics[width=1\linewidth]{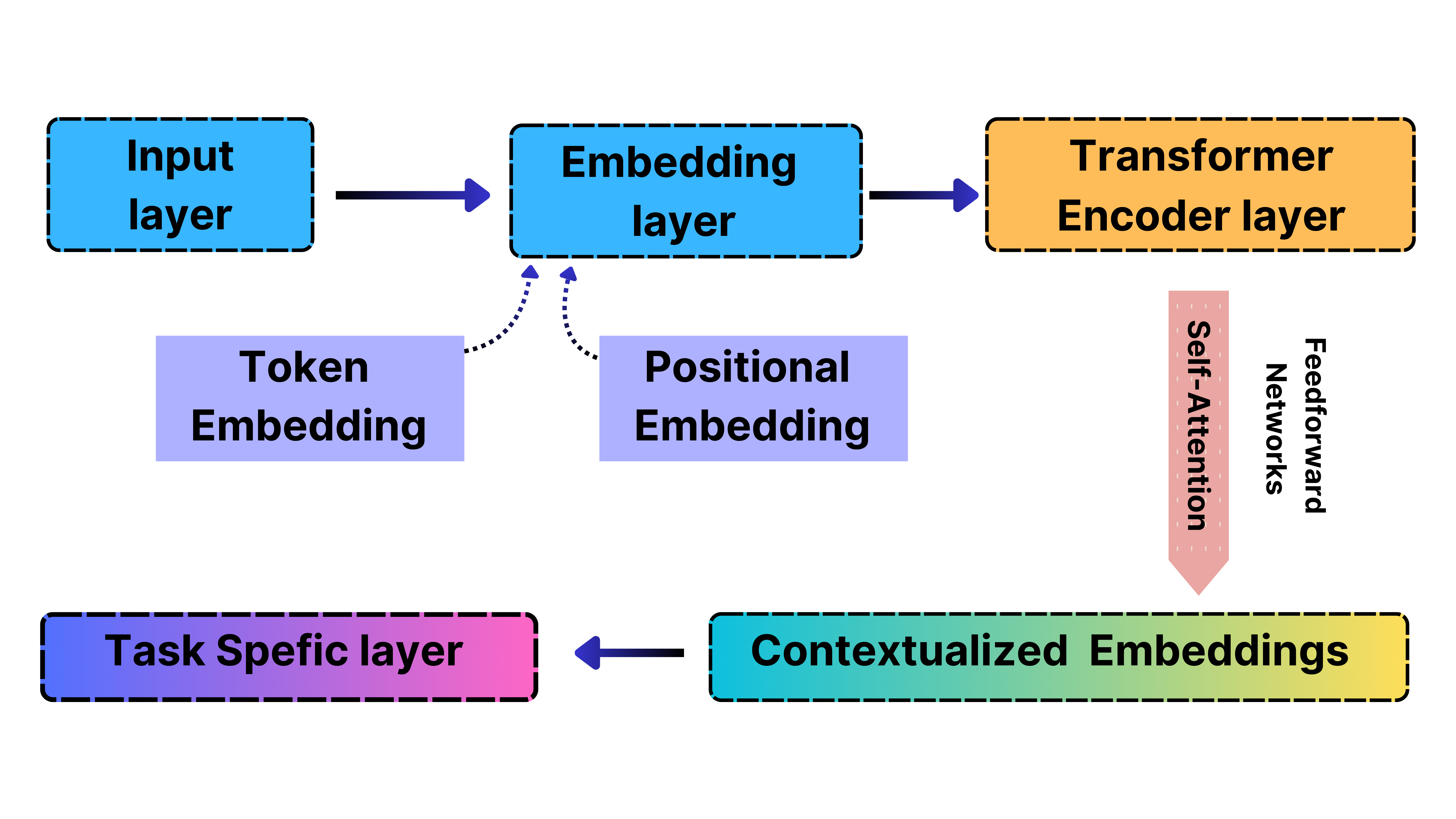}
     \vspace{-3em}
    \caption{Overview of CodeBERT strategies, illustrating the preprocessing steps and transformer encoder layers.}

    \label{fig:f2}
\end{figure}

Based on BERT procedures, there are 12 transformer encoder layers in CodeBERT. Each of these layers applies a self-attention mechanism and a feed-forward network to convert the input embeddings into contextual representations. The self-attention mechanism sheds light on the connections between tokens, while the feed-forward network modifies non-linearity. In this study, the process begins with an input layer that receives a series of tokens of programming code. Each token pass through token embedding that transforms the tokens into a high-dimensional vector that encapsulates its semantic meaning. Addition of positional embeddings provides sequence information and position of tokens, due to the transformers models lack an inherent understanding of token order. Afterwards, the output then passed into CodeBERT layers that produces contextualized embeddings and grasp the nuances of input code. Subscequently, these embeddings are then fed into a task-specific layer such as our motivations vulnerability detection. Figure \ref{fig:f2} depicts the strategies applied in our study.

\subsection{Models Description }

LLMs have revolutionized the field of NPD identification and are becoming a game-changer for analyzing code snippets. These advancements in LLMs provide deeper insights with the utilization of advanced machine learning techniques. In this study, we went through various advanced machine-learning techniques and found that the DeLLNeuN model outperformed all others. To enhance the vulnerability prediction performance, we propose implementing a tailored neural network model that incorporates several key architectural elements includes dropout layers, dense layers, and ReLU activation functions. Finally, the sigmoid function transformed the prediction into probabilities to make the model suitable for the binary classification tasks. The following section outlines the working procedures of each applied model and highlights all applied hyperparameters in Table \ref{tab:models_hyperparameters}.

\subsubsection{CodeBERT} 
CodeBERT \cite{ref21} is a pre-trained language model specifically designed for tasks involving source code and natural language. The model is based on the BERT (Bidirectional Encoder Representations from Transformers) architecture and includes 125 million parameters. CodeBERT uses a Transformer-based neural architecture and is trained with a hybrid objective function. Additionally, CodeBERT analyzes context in both directions (left to right and right to left), allowing it to fully understand the meaning of the text. It works by encoding snippets of source code and natural language into dense vector representations that capture their underlying semantic meaning.

\subsubsection{GraphCodeBERT} 
GraphCodeBERT \cite{ref22} is a pre-trained model built specifically for programming languages, offering an alternative to traditional syntax-based structures like abstract syntax trees (AST). In this model, the code is represented as a graph where nodes correspond to variables, and edges show the relationships that trace where values originate between these variables. This setup helps to map out where each value comes from and how different variables are connected. The model incorporates a graph-guided masked attention function to efficiently integrate code structure, relying on the transformer neural architecture as its core. The model is initially trained on a vast collection of source code and then fine-tuned for specific tasks such as summarizing code, searching through codebases, and detecting vulnerabilities.

\subsubsection{RoBERTa}
This is an enhanced version of the BERT model, featuring 340 million parameters and designed to improve performance on a range of natural language understanding tasks. Unlike BERT, RoBERTa \cite{ref23} is trained on a much larger dataset and handles longer text sequences, while omitting the Next Sentence Prediction (NSP) task. This adjustment allows RoBERTa to capture more nuanced language patterns and context. As a result, it sets new standards in performance on several benchmarks, including GLUE, RACE, and SQuAD \cite{ref24}.

\begin{table}[H]
\centering
\caption{Overview of Applied Models and their Hyperparameters.}
\label{tab:models_hyperparameters}
\resizebox{0.5\textwidth}{!}{
\begin{tabular}{|l|l|}
\hline
\textbf{Models name} & \textbf{Parameters} \\ \hline
CodeBERT & \begin{tabular}[c]{@{}l@{}}evaluation\_strategy = "epoch",\\  num\_train\_epochs = 3,\\ weight\_decay = 0.01,\\ transformer\_layers = 12\end{tabular} \\ \hline
DeLLNeuN & \begin{tabular}[c]{@{}l@{}}max\_length = 512,\\ batch\_size = 8,\\ Padding = "True",\\ Truncation = "True",\\ return\_attention\_mask = "True",\\ return\_tensors = "pt",\\ drop\_out = 0.3,\\ dense\_output dimension = 512,\\ logistic\_function = sigmoid,\\
activation\_function = ReLU,\\ learning rate: 2e-5,\\ optimizer = AdamW(weight\_decay=0.01)\end{tabular} \\ \hline
GraphBERT & \begin{tabular}[c]{@{}l@{}}Padding = "True",\\ Truncation = "True",\\ max\_length = 512,\\ num\_train\_epochs = 3,\\ weight\_decay = 0.01\end{tabular} \\ \hline
RoBERTa & \begin{tabular}[c]{@{}l@{}}truncation = True,\\ return\_tensors='pt',\\ Padding = “max length” (512),\\ \end{tabular} \\ \hline
LSTM & \begin{tabular}[c]{@{}l@{}}max\_vocab\_size = 10000,\\ max\_sequence\_length = 500,\\ embedding\_dim = 100,\\ units = (64, 32),\\ activation = 'relu'\end{tabular} \\ \hline
GPT2 & \begin{tabular}[c]{@{}l@{}}evaluation\_strategy = "steps",\\ save\_steps=500\end{tabular} \\ \hline
\end{tabular}
}
\end{table}

\subsubsection{LSTM}
LSTM \cite{ref25} stands for (Long Short Term Memory), networks are a specialized type of recurrent neural network (RNN) designed to process and analyze sequences of data. The ability of LSTMs to preserve long-term dependencies within the data makes the model unique. However, LSTM achieves this through memory cells and three key gates: forget, input, and output gates. All of these key components work together to control the flow of information. The memory cell stores important information over time, while the gates decide which information to keep, update, or discard. This enables LSTMs to effectively capture and utilize relevant patterns in the data, making them particularly useful for complex sequential tasks.

\subsubsection{GPT-2}
GPT-2 \cite{ref26} is a powerful language model developed by OpenAI, designed to generate human-like text based on the input it receives. GPT-2 is a transformer-based model that was pre-trained on a massive dataset of 8 million web pages. Its core strength lies in its ability to anticipate the next word in a sentence, which enables it to generate contextually relevant text. This predictive capability makes GPT-2 particularly effective when fine-tuned for specific tasks, like working with code snippets, where it can adapt to provide accurate and relevant outputs.

\subsection{Proposed Model (DeLLNeuN)}

CodeBERT is a bi-modal pre-trained model designed to handle both programming languages and natural language, and it has been effectively applied to various tasks involving source code. Typically, the output from CodeBERT's final layer is used as a representation of code semantics for fine-tuning tasks \cite{ref27}. However, the pre-trained model CodeBERT utilized as feature embedders to transform these textual values into numerical values and uncover the relevant contextual information \cite{ref28,ref29}. Additionally, this approach might overlook valuable information that could be captured by the other layers of CodeBERT. To more effectively utilize the information each layer of CodeBERT can provide, we introduce a new approach called DeLLNeuN. Instead of just using the final layer’s output, DeLLNeuN defines multiple layers on top of CodeBERT. Our method extracts and utilizes the information from all layers of CodeBERT, treating this as a sequence of representations to enhance the fine-tuning process for source code-related tasks.

DeLLNeuN is designed to handle binary classification tasks using features extracted from CodeBERT. After CodeBERT generates a contextualized embedding of the input code, DeLLNeuN processes this embedding through several neural network layers to make a classification decision. At first, a dropout layer, set to 30\%, randomly deactivates some neurons during training. This technique helps prevent overfitting by ensuring the model doesn't become too dependent on any single neuron.

\begin{figure}
    \centering
    \includegraphics[width=1\linewidth]{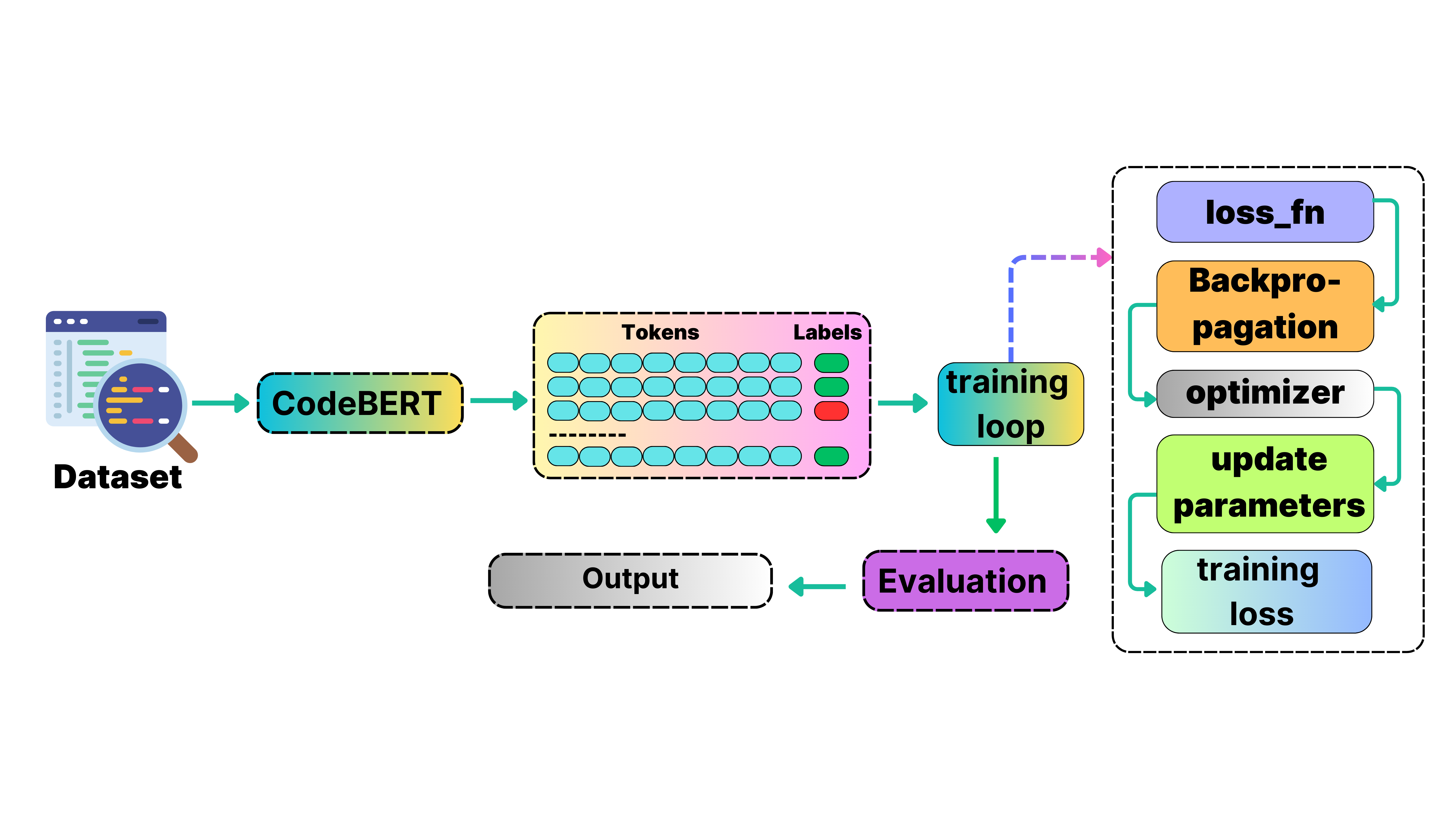}
    \vspace{-4em}
    \caption{Architecture of the DeLLNeuN model, showing the integration of multiple layers on top of CodeBERT.}
    \label{fig:f4}
\end{figure}

Sequentially, the model uses a dense layer with 512 units to reduce the dimensionality of the feature space. A ReLU activation function is applied here to introduce non-linearity, allowing the model to recognize more complex patterns, as illustrated in Figure \ref{fig:f4}. 

Finally, the model passes through a classifier layer, which is another fully connected dense layer. This layer produces logits for each of the two classes (vulnerable or non-vulnerable). These logits are then transformed into probabilities using a sigmoid function, indicating whether the input code is vulnerable or non-vulnerable.

\subsection{Evaluation Metrics}

To validate our proposed model, we utilize several validation metrics such as precision, sensitivity (recall), F1-score, and accuracy. These metrics offer an in-depth understanding of the model’s performance in identifying vulnerabilities in source code \cite{ref30, ref31}. We considered these metrics because accuracy is defined as the ratio of accurately predicted instances to total instances. However, the accuracy metric alone is not sufficient to differentiate between the types of errors, such as false positives and false negatives, which are crucial in this study.

\textbf{Precision} stands for the ratio of accurately predicted vulnerabilities out of all the positive predictions made by the model. \textbf{Sensitivity} evaluates the proportion of true vulnerability cases that the model accurately identified, with a focus on reducing false non-vulnerabilities. By combining sensitivity and precision into a single metric, the \textbf{F1-score} provides an unbiased assessment of the model's performance. The mathematical equations for the applied metrics are shown below.

\begin{equation}
\text{ReLU}(x) = \max(0, x) \label{eq:relu}
\end{equation}
Here, \(x\) represents any input; the ReLU function returns \(x\) if \(x\) is greater than 0, and 0 otherwise.

\begin{equation}
\text{Recall} = \frac{TP}{TP + FN} \label{eq:recall}
\end{equation}

\begin{equation}
\text{Precision} = \frac{TP}{TP + FP} \label{eq:precision}
\end{equation}

\begin{equation}
\text{F1-score} = 2 \times \frac{\text{Precision} \times \text{Recall}}{\text{Precision} + \text{Recall}} \label{eq:f1}
\end{equation}

Here, true negative (TN) indicates that the model accurately identifies a non-vulnerable case, whereas a true positive (TP) indicates that the model correctly identifies a vulnerable case. However, if the model incorrectly classifies a non-vulnerable case as vulnerable, that falls into a false positive (FP). In contrast, a false negative (FN) refers to when the model attempts to identify a vulnerability and classifies it as non-vulnerable.

\section{Experimental Result}

In this study, we used several methods to develop an effective predictive outcome for identifying vulnerabilities in source code. The outcomes of these models are shown in Table \ref{tab:experimental_results}, and the assessment of our suggested model is provided in Table \ref{tab:Table 5}.

\begin{table}[htbp]
\centering
\caption{Experimental analysis results with the Draper VDISC dataset.}
\label{tab:experimental_results}
\resizebox{0.75\linewidth}{!}{%
\begin{tabular}{|l|c|c|c|c|}
\hline
\textbf{Models} & \textbf{Accuracy} & \textbf{Precision} & \textbf{F1-Score} & \textbf{Recall} \\ \hline
CodeBERT  & 0.76 & 0.77 & 0.76 & 0.76 \\ \hline
GraphBERT & 0.76 & 0.76 & 0.75 & 0.76 \\ \hline
RoBERTa   & 0.76 & 0.72 & 0.78 & 0.85 \\ \hline
LSTM      & 0.71 & 0.71 & 0.71 & 0.71 \\ \hline
GPT-2     & 0.73 & 0.71 & 0.73 & 0.76 \\ \hline
\end{tabular}%
}
\end{table}

The findings highlighted in Table \ref{tab:experimental_results}, a few significant differences across the models. RoBERTa, having the highest recall performance meaning particularly effective at identifying vulnerabilities, though its precision was a bit lower. According to this, RoBERTa was quite good at spotting potential vulnerability, although it also flagged a few false positives. The accuracy and recall of both CodeBERT and GraphBERT approaches were consistently 0.76. While CodeBERT somewhat outperformed GraphBERT in terms of precision and F1-Score. Besides, GPT-2 models performed well, demonstrating a balanced recall and precision, while falling short of RoBERTa's recall.

\begin{table}[htbp]
\centering
\caption{Experimental outcomes of DeLLNeuN model.}
\vspace{1em}
\label{tab:Table 5}
\resizebox{\linewidth}{!}{%
\begin{tabular}{|c|c|c|c|c|c|c|}
\hline
\textbf{Model name} & \textbf{Process} & \textbf{Steps} & \textbf{Accuracy} & \textbf{Precision} & \textbf{F1-Score} & \textbf{Recall} \\ \hline
\multirow{3}{*}{\textbf{DeLLNeuN}} & \multirow{2}{*}{Direct} & 1 & 0.861 & 0.858 & 0.859 & 0.863 \\ \cline{3-7} 
 &  & 2 & 0.864 & 0.871 & 0.865 & 0.872 \\ \cline{3-7} 
 &  & \textbf{3} & \textbf{0.871} & \textbf{0.881} & \textbf{0.875} & \textbf{0.877} \\ \cline{2-7} 
 & \multirow{5}{*}{Cross validation} & 1 & 0.863 & 0.872 & 0.871 & 0.870 \\ \cline{3-7} 
 &  & 2 & 0.862 & 0.872 & 0.861 & 0.863 \\ \cline{3-7} 
 &  & 3 & 0.859 & 0.861 & 0.861 & 0.860 \\ \cline{3-7} 
 &  & \textbf{4} & \textbf{0.872} & \textbf{0.873} & \textbf{0.874} & \textbf{0.872} \\ \cline{3-7} 
 &  & 5 & 0.864 & 0.862 & 0.860 & 0.865 \\ \hline
\end{tabular}%
}
\end{table}

The results in Table \ref{tab:Table 5} provide a detailed overview of DeLLNeuN’s performance across different processes and steps. In the direct process, DeLLNeuN delivered impressive outcomes with accuracy ranging from 0.861 to 0.871. The third step displayed the best results, with 0.871 accuracy, 0.881 precision, and 0.877 recall. This indicates that DeLLNeuN performed consistently well and its precision improved with time. 

In cross-validation, the model performed well with accuracy ranging from 0.859 to 0.872. With an accuracy and precision of 0.872 and 0.873, respectively, the fourth cross-validation achieved the highest scores, indicating its stability and reliability.

Table \ref{tab:Table 5} showcasing a detailed overview of DeLLNeuN’s performance across different processes and steps. In the direct process, DeLLNeuN delivered impressive outcomes with accuracy ranging from 0.861 to 0.871. The third step displayed the best results, with 0.871 accuracy, 0.881 precision, and 0.877 recall. This indicates that DeLLNeuN performed consistently well and its precision improved with time. In cross-validation, the model performed well with the accuracy ranging from 0.859 to 0.872. With an accuracy and precision of 0.872 and 0.873, respectively, the fourth cross-validation achieved the highest scores, indicating its stability and reliability.

In general, DeLLNeuN performs better than the earlier models, particularly in terms of precision and recall. DeLLNeuN consistently produced better precision and F1-Score, indicating a more balanced performance overall, while RoBERTa had a more robust recall than the other models. The cross-validation results also reflect a stable performance, though there were some fluctuations but it is normal in such evaluations. In comparison to the previous models, DeLLNeuN's performance is more constant and dependable, indicating that it is a strong contender in vulnerability detection.

\section{Limitations for the Future Study}

In this research, our key objective was to implement an AI to achieve high efficiency in identifying vulnerabilities (CWE-476) from source code, and our proposed model shows promising results. However, there are a few areas for improvement in future research. Due to resource constraints, we only utilized a portion of the main dataset. Additionally, the model was trained on a single GPU. In future work, expanding computational resources would allow for more robust training and improved accuracy in detecting a wider range of vulnerabilities.

\section{Conclusion}

In conclusions, this study demonstrates the potential uses of LLM’s model specially CodeBERT for effectively identify vulnerabilities like CWE-476 in source code. Through the use of CodeBERT, the model is able to effectively identify vulnerabilities by utilizing deep learning’s, to better understanding of code semantics. This capacity is further enhanced by its layered architecture, which incorporates several specialized neural network layers to further hone the detecting process. This enables the model to pinpoint subtle vulnerabilities that conventional static analysis methods could not detect. The proposed model’s accuracy and consistency, suggesting that it has the potential to be useful tool in software security. This breakthrough opens new possibilities for AI driven tools for vulnerability detentions, paving the way for a more secure digital realm.

\section*{Conflict of Interest}
The authors declare that they have no conflict of interest.
They also declare that they do not have any competing financial interests.

\bibliographystyle{ieeetr}

\end{document}